\documentclass[journal]{IEEEtran}

\ifCLASSOPTIONcompsoc
  \usepackage[nocompress]{cite}
\else
  \usepackage{cite}
\fi

\ifCLASSINFOpdf
   \usepackage[pdftex]{graphicx}
\else
\fi

\usepackage{amsmath}
\usepackage{amssymb}
\usepackage{algorithmicx}

\usepackage{array}

\ifCLASSOPTIONcompsoc
  \usepackage[caption=false,font=footnotesize,labelfont=sf,textfont=sf]{subfig}
\else
  \usepackage[caption=false,font=footnotesize]{subfig}
\fi

\usepackage{fixltx2e}

\usepackage{stfloats}

\usepackage{url}
\usepackage[T1]{fontenc}

\begin{document}

\title{On the separation of shape and temporal patterns in time series\\-Application to signature authentication-}

\author{Pierre-Fran\c{c}ois Marteau,~\IEEEmembership{}
\IEEEcompsocitemizethanks{\IEEEcompsocthanksitem  M. Marteau was with the 	Institut de Recherche en Informatique et Syst\`{e}mes Aléatoires (IRISA), Universit\'{e} Bretagne Sud, Vannes, France.\protect\\
E-mail: see http://people.irisa.fr/Pierre-Francois.Marteau/}
\thanks{}}

\markboth{Journal of \LaTeX\ Class Files,~Vol.~xx, No.~8, MMM~AAA}%
{Shell \MakeLowercase{\textit{et al.}}: Bare Demo of IEEEtran.cls for Computer Society Journals}

\IEEEtitleabstractindextext{%
\begin{abstract}

In this article we address the problem of separation of shape and time components in time series. The concept of shape that we tackle is termed temporally \textit{neutral} to consider that it may possibly exist outside of any temporal specification, as it is the case for a geometric form. We propose to exploit and adapt a probabilistic temporal alignment algorithm, initially designed to estimate the centroid of a set of time series, to build some heuristic elements of solution to this separation problem. We show on some controlled synthetic data that this algorithm meets empirically our initial objectives. We finally evaluate it on real data, in the context of some on-line handwritten signature authentication benchmarks. On the three evaluated tasks, our approach based on the separation of signature shape and associated temporal patterns is positioned slightly above the current state of the art demonstrating the applicative benefit of this separating problem.
\end{abstract}

\begin{IEEEkeywords}
Time series averaging, Shape and Time Separation, Time Elatic Kernels, Handwritten Signature Authentication.
\end{IEEEkeywords}}

\maketitle

\IEEEdisplaynontitleabstractindextext

\IEEEpeerreviewmaketitle

\section{Introduction}\label{sec:introduction}

\IEEEPARstart{S}{ome} time series are quite specific in a way that they describe shapes. This is the case, for instance, for a dynamical system reaching a limit cycle in its observable state space. 
The concept of shape could have, however several meanings, and consequently a shape can be characterized through different perspectives. With a mathematical point of view, geometrical shapes can be defined as the solution set of an ensemble of equations. Simple shapes such as ellipses or complex structure such as strange attractors or fractal sets are fully entering into this kind of conceptual definition for a shape. In such view, if the equations defining the shape are time independent, the shape is fully defined as a geometric object that exist outside any temporal specification. Shapes, in another hand, can also be described indirectly through parametric equations that provide multivariate time series somehow forming the contour of a shape. The parameter entering into such equations is generally, by convention, referred to as the time variable.  Apart from such purely mathematical formalizations, shapes have been also early associated to visual patterns. This association has motivated considerable research work in machine vision to identify or recognize shape-object \cite{Andreopoulos2013}, from their contour, texture, colors, etc. Progressively, from this somehow restricted view that assimilates a shape to an identifiable visual stimuli, a wider range of shape characterization associated to other perceptual stimuli has emerged. The acoustic channel  has thereby led to the definition of acoustic patterns, i.e. phonemes, characteristic noises, musical forms, etc., that share a lot of similarity with the visual shapes, at least in their definitions. Thereby, major works have been done, in particular in the speech recognition area, to model and identify shapes "encoded" in acoustic time series.  Beyond the acoustic channel modelization, time series analysis has allows for the characterization of very general "temporal" shapes \cite{NIENNATTRAKUL2012}, just as images or videos processing has allowed for visual-shapes. Today, temporal, visual or visuo-temporal shape analysis find applications in any area of our digitized world.\\

We adopt in this article a more restrictive definition for shape, somehow closer to a geometrical or mathematical characterization, according to which a form can exist outside any temporal consideration. Let us consider formally the ellipse example that is completely defined by the following equation \ref{eq:ellipse}.
	\begin{equation}
	\frac{x^2}{a^2}+\frac{y^2}{b^2}=1
	\label{eq:ellipse}
	\end{equation} 
The set of solutions in this equation completely and \textit{atemporaly} defines an ellipse. However, as stated above, this shape can also be defined through a set of parametric equations (Eq. \ref{eq:ellipseParam}) that reintroduce a parameter, $t$, which we call the temporal variable.	
	\begin{equation}
	\left\{
	\begin{array}{ll}
	x_t=a.cos(\omega t) \\
	y_t=b.sin(\omega t)
	\end{array}
	\right.
	\label{eq:ellipseParam}
	\end{equation}
	
In this second formulation, the time series $x(t)$ and $y(t)$ also completely define the ellipse shape. These function of $t$ can be understood as access functions to the elements of the set of solutions. 

When this parameter $t$ is sampled, it can indeed be considered as an index allowing to enumerate the pairs $(x_{t_n}, y_{t_n})$ that are solutions of the preceding equations and regrouped this time in a multiset $\mathcal{E} = \{(x_ {t_n}, y_ {t_n})\}_{n = 0,1, \ cdots}$, with $t_n = n/f_e$ and $f_e$ the sampling frequency. We can then assimilate the shape associated with the ellipse to the multiset $\mathcal{E}$.

Shapes that can be characterized in this way are relatively diverse and potentially complex. They emerge in particular in the asymptotic behaviors of some nonlinear dynamic systems, stationary or not, described by differential  or difference equations, in the form of limit cycles or strange attractors like the fractal attractors of Rosler or Lorentz. We can suppose that these shapes may also emerge in some traces produced by nonstationary stochastic generative systems, but in a noisy or hidden way, since the shapes are in this case accessible through an observable which depends on the time variable.\\

Based on this notion of \textit{atemporal} shape, we are interested in the problem of the extraction of the shape (that is assumed to be \textit{atemporal}) and associated temporal functions, given a subset of time series assumed to be produced by a stochastic generative process. This problem has been addressed in terms of the characterization of amplitude and phase variabilities. In particular the problem of separation of
amplitude and phase variation has been studied for the case where one observes multiple realisations of random point processes \cite{Panaretos2016}. For time series clustering, constrained optimization solutions have also been proposed in \cite{Sangalli2010}. In this paper we tackle this separation problem through a probabilistic interpretation of time elsatic kernel \cite{Marteau2019} which leads to propose an efficient algorithm for decoupling phase and shape in discrete time series subsets.

In the second part of this paper, we formalize this problem of shape/time separation for a restrictive category of stochastic processes expressed in discrete time and detail the heuristic approach that we develop to solve this problem, conjecturing that it is NP-complete. The third part is devoted to an experiment which allows to evaluate initially the algorithmic solution developed on synthetic data that we fully control. In a second step we test our algorithm on real data as part of on-line signature authentication tasks. In the last conclusive section, we discuss our findings and the prospects that they suggest.\\

\section{Problem and proposed solution formalization}

\subsection{Context}
\label{sec:context}
We consider the following $\mathcal{G}$ class of stochastic generative processes that can be formalized in a parametric form:
\begin{equation}
o(t) = G(\xi(t))+\epsilon_\sigma(t)
\label{eq:processus}
\end{equation}

with $t \in \{0,1,2, \cdots\}$ the discrete time variable, $\xi \in \Xi$ a function, possibly stochastic, monotonously increasing with $t$, $G \in \Gamma$ a nonlinear function randomly drawn from $\Gamma$,  $o(t)$ the observable output of the process at timestamps $t$ and $\epsilon$ a Gaussian noise with zero mean and standard deviation $\sigma$. We will denote such process $(\Gamma,\Xi, \epsilon_\sigma)$.  

The set $\Gamma$ will characterize what we call the shape variability and the set $\Xi$ the temporal variability. \\

Thus, if we observe $N$ distinct realizations of the previous generative process, we get a set of $N$ traces $\{o_i(t)\}_{i=1,\cdots,N}$ such that:

\begin{eqnarray}
\forall i, o_i(t) = G_i(\xi_i(t))+\epsilon_i(t)
\label{eq:processus}
\end{eqnarray}
with  $G_i \in \Gamma$, $\xi_i \in \Xi$ and $\epsilon_i$ a Gaussian noise.
\\

In the absence of a strictly atemporal shape, we will call (temporally) \textit{neutral} shape associated to the process defined by Eq.\ref{eq:processus} 
the time series $O_s(t) = E(G(E(\xi(t))))$, where $E$ represents the mathematical expectation, and $t \in \{0,1,2, \cdots\}$. For the \textit{neutral} shape, the associated temporal function $ \xi_s(t)$ corresponds to the mathematical expectation of observed $\xi$ functions.\\\\

As an example, the stochastic process defined by
\begin{enumerate}
\item matrix  $G_{a,b}$ defined in Eq. \ref{eq:ellipse_theta},
\item $\xi_\omega(t) = \omega \cdot t$,
\item constraints on the parameters, $a=a_0\cdot(1+U_{[-.5,.5]})$,
$b=b_0\cdot(1+U_{[-.5,.5]})$, \\$\omega=\omega_0\cdot(1+U_{[-.1,.1]})$, $\varphi=\varphi_0+U_{[-.1,.1]}$  where $U_{[u,v]}$ is the uniform law defined on interval $[u,v]$, and
\item the additive Gaussian noise $\epsilon_\sigma(t)$, 
\end{enumerate}
belongs to the $\mathcal{G}$ class of process, sets $\Gamma$ and $\Xi$ being respectively defined by the constraints on $(a,b)$ and $(\omega,\varphi)$ parameters. 

\begin{equation}
G_{a,b}(t)=
\begin{bmatrix}
a \cdot cos(\xi_\omega(t)+\varphi(t)) & 0 \\
0 & b \cdot sin(\xi_\omega(t)+\varphi(t))
\end{bmatrix}
\label{eq:ellipse_theta}
\end{equation}

Here, $\Gamma=\{G_{a,b} | a=a_0\cdot(1+U_{[-.25,.25]}), b=b_0\cdot(1+U_{[-.25,.25]})\}$ and $\Xi=\{\omega t +\varphi | \omega=\omega_0\cdot(1+U_{[-.1,.1]}), \varphi=\varphi_0+U_{[-.25,.25]}\}$.\\

The \textit{neutral} shape for this process is $O_s(t) = G_{a_0,b_0}(\omega_0 t + \varphi_0)$.\\

\subsection{Statement of the separability problem}

Given  $\{o_i(t)\}_{i=1,2,\cdots,N}$, $N$ observations assumed to be produced by a stochastic process class $\mathcal{G}$,\\

\hspace{10mm}\textbf{Can we estimate} $\Gamma$, $\Xi$, $O_s$ and $\{\xi_i(t)\}_{i=1,2,\cdots,N}$ under the conditions that $\forall i$, $o_i(t) = G_i(\xi_{i}(t))+\epsilon_i(t)$, $G_i \in \Gamma$, $\xi_i \in \Xi$ and $\epsilon_i$ a Gaussian noise with zero mean?\\

Answering this question is to separate the \textit{neutral} shape from the temporal functions. This separation makes it possible to consider the study of shape variability independently of temporal variability.

\subsection{Elements of heuristic solution to the separability problem}

The general problem of separability as previously stated is under-constrained (or ill-posed) and, as such, probably does not admit a solution in the general case. On the other hand, we empirically show that the algorithm proposed in \cite{Marteau2019} that estimates an average time series of a set of time series using temporally elastic kernel provides a way to approximate the concept of \textit{neutral} shape. Moreover, the probabilistic formalization of the temporal alignment mechanism that is suggested is particularly suited to address our separation problem.

We present below the general concepts of this algorithm (for details, see \cite{Marteau2019}) and show how it offers elements of answer to the previously stated problem of separability.\\

Unlike the dynamic time warping algorithm (DTW) \cite{Sakoe71} which considers only the best alignment path to match pairs of time series, the time elastic kernels \cite{CuturiVert2007, MarteauGibet2014} consider the sum of the alignment scores of all possible matching paths. The score assigned to a path is the result of the product of the local matching kernels evaluated along the path.

Let $o_1^n = o_1 o_2 \cdots o_n$ and ${o'}_1^{n'} = o'_1 o'_2 \cdots  o'_{n'}$ be two time series (observations) with respective lengths $n$ and $n'$. The local kernel that quantifies the proximity of two samples $o_i$ and $o'_j$ is in general an exponential kernel : $k(o_i,o'_j)= e^{-\nu\cdot ||o_i,o'_j||^2}$. By interpreting this local kernel as a local probability function \cite{Marteau2019}, it is possible to consider the alignment of two time series as the mathematical expectation of the outputs generated by a stochastic finite state machinery.

The considered stochastic \textit{automata} is a Hidden Markov process defined on a finite set of hidden states $\mathcal{S} =\{ S_{1,1},S_{1,2}, \cdots, S_{n,n'}, S_{n+1,n'+1}\}$. Each state $S_{i,j}$ characterizes the alignment between two samples $o_i$ and $o'_j$, $S_{1,1}$ is assimilated to the initial state and $S_{n,n'}$ as the final state. \\

The hidden variable (at step $\tau$) $z_\tau$ takes value in $\mathcal{S}$. The index $\tau$ follows one of the admissible alignment paths and takes value in  $\{(1,1), ...,(n,n')\}$. To take into account monotonic alignment constraints (as in DTW), we impose  $\tau=(i,j)$, $\tau+1 \in \{(i,j+1), (i+1,j), (i+1,j+1)\}$ and $\tau-1 \in \{(i,j-1), (i-1,j), (i-1,j-1)\}$.\\

The \textit{a posteriori} probability for the process to be in state $S_{i,j}$ at step $\tau$, given the observed sequences $o_1^n$ et ${o'}_1^{n'}$ est $P(z_\tau=S_{i,j}|o_1^n,{o'}_1^{n'},\theta)$, where $\theta$ is the vector of parameters of the Hidden Markov Model. This \textit{a posteriori} probability is estimated through forward and backward recursions, $\alpha$ and $\beta$ defined as follows\\

$\alpha_{\tau}(t,t')= P(o_1^t,{o'}_{1}^{t'},z_{\tau}=S_{t,t'}|\theta)$ is the probability for the process to be at state $S_{t,t'}$ at step $\tau$ once the partial observation sequences $(o_1^t, {o'}_{1}^{t'})$ are aligned.\\
\indent $\beta_{\tau}(t,t')=P(o_t^n,{o'}_{t'}^{n'}|z_\tau=S_{t,t'},\theta)$ is the probability of the alignment of the pair of partial sequences $(o_t^n, {o'}_{t'}^{n'})$ given the alignment process is at state $S_{t,t'}$ at step $\tau$. \\

The matching of any pair of samples $o_i$ and $o'_j$ can then be interpreted as the probability of presence of the stochastic alignment \textit{automata} on cell $S_{i,j}$.

The marginal probability for the process to be at step $\tau$ in one of the states of the subset $S_{t,\bullet} = \{S_{t, 1}, S_{t, 2}, \cdots, S_{t, n'}\}$, being given the observations $o_{1}^{n}$ and ${o'}_{1}^{n'}$, (that is, the probability for the process to visit at step $\tau$ one of the state of $S_{t,\bullet}$, or $o_t$ to be aligned with one of the samples in the series ${o'}_1^{n'}$) is given by:

\begin{equation}
\begin{array}{ll}
P(z_\tau \in S_{t,\bullet}|o_1^n,{o'}_{1}^{n'},\theta)=\sum_{t'}P(z_\tau=S_{t,t'}|o_1^n,{o'}_{1}^{n'},\theta)
\end{array}
\end{equation}

\noindent
For all $t$ and $t'$, the conditional probability of visiting the state $S_{t, t'}$ knowing the two series of observations, the parameter $\theta$ and the subset $S_{ t, \bullet}$, in other words the probability that $o_t$ and $o'_{t'}$ are aligned given that $o_t$ is aligned with one of the samples of the series ${o'}_{ 1}^{n'}$ is:
\begin{equation}
\label{CondProp1}
\begin{array}{ll}
P(z_\tau=S_{t,t'}|o_1^n,{o'}_{1}^{n'}, z_\tau \in S_{t,\bullet}, \theta) = \\
\hspace{5mm}\frac{P(z_\tau=S_{t,t'}|o_1^n,{o'}_{1}^{n'},\theta)}{P(z_\tau \in S_{t,\bullet}|o_1^n,{o'}_1^{n'},\theta) }
\end{array}
\end{equation} 


The mathematical expectation of alignments of samples in the ${o'}_{1}^{n'}$ series with the sample $o_t$ (given that $o_t$ is aligned by the automaton) and similarly the expectation of the time occurrences associated with the samples of the series ${o'}_{1}^{n'}$ aligned with $o_t$ are defined by:


\begin{equation}
\label{Expectation1} 
\begin{array}{ll}
E(o'|o_t) \propto \sum\limits_{t'=1}^{n'} {o'}_{t'}  P(z_\tau = S_{t,t'}|o_1^n,{o'}_1^{n'},z_\tau \in S_{t,\bullet},\theta)\\
E(t'|o_t) \propto \sum\limits_{t'=1}^{n'} {t'}  P(z_\tau = S_{t,t'}|o_1^n,{o'}_1^{n'},z_\tau \in S_{t,\bullet},\theta)\\
\end{array}
\end{equation} 

The previous mathematical expectations (Eq. \ref {Expectation1}) are the basis of the proposed procedure for estimating a time elastic centroid for a set of time series.

Let $\mathcal{O}=\{{}^k\! o_1^{n_k}\}_{k=1 \cdots N}$ be a subset of time series, $r_1^{n}$ a reference time series ($r_1^{n}$ can be initialized with the medoid of $\mathcal{O}$).
The centroid estimate of $\mathcal{O}$ is defined by the tuple $(O_s, \xi_s)$ where $O_s$ is a time series of length $n$ and $\xi_s$ is the sequence of time occurrences associated to the sample of $O_s$


\begin{equation}
\label{eq:TEKA}
\begin{array}{ll}
O_s(t)=\frac{1}{N} \sum\limits_{k=1}^N E({}^k\! o|r_t)\\
\hspace{5mm} \propto \frac{1}{N} \sum\limits_{k=1}^N \sum\limits_{{}^k\! t=1}^{n_k}   {}^k\! o_{{}^k\! t} P(z_\tau = S_{t,{}^k\! t}|r_1^{n},{}^k\! o_1^{n_k}, z_\tau \in S_{t,\bullet}, \theta)\\
\xi_s(t)=\frac{1}{N} \sum\limits_{k=1}^N E({}^k\! t|r_t)\\
\hspace{5mm} \propto\frac{1}{N} \sum\limits_{k=1}^N \sum\limits_{{}^k\! t=1}^{n_k}   {}^k\! t P(z_\tau = S_{t,{}^k\! t}|r_1^{n},{}^k\! o_1^{n_k}, z_\tau \in S_{t,\bullet}, \theta)\\
\end{array}
\end{equation} 

$(O_s,\xi_s)$ is in general a non-uniformly sampled time series: $\xi_s(t)$ is the mathematical expectation of the time of occurrence for sample  $O_s(t)$. If necessary, a uniform re-sampling can straightforwardly be used to get back to a uniformly sampled time series.\\

By successive iterations, i.e. by replacing the reference time series $r_1^{n}$ by the new centroid estimates $O_s$ evaluated at the end of each iteration, until no more improvement can be obtained (decrease in inertia), we get a final estimate of the centroid for $\mathcal{O}$. \\

It should be noted here that the underlying multiple alignment problem being NP-complete \cite{WangJ1994} and characterized with an exponential complexity \cite{Just99} with the size of $\mathcal{O}$, only sub-optimal heuristics allow to approach this notion of temporally elastic centroid for sets $\mathcal{O}$ of more than tens of time series. \\


We conjecture that, at the end of the iterative process, $(O_s, \xi_s)$ as defined by the equations Eq.\ref{eq:TEKA} is an empirical estimator for the \textit{neutral} shape associated to the set $\mathcal{O}$. Furthermore, equations \ref{Expectation1} also provide access to temporal function estimators $\xi_i$ associated with the elements of $\mathcal{O}$. Thus, for $o_i = {o_i}_1^{n_i} \in \mathcal{O}$, and the \textit{neutral} shape $O_s$ estimated on $O$, the estimate of the temporal function $\xi_i$ associated with  $o_i$ is:

\begin{equation}
\label{eq:xi} 
\begin{array}{ll}
\tilde{\xi}_i (t) = E(t'|C_t) \\
\hspace{8mm}\propto \sum\limits_{t'=1}^{n_i} {t'}  P(z_\tau = S_{t,t'}|{O_s}_1^n,{o_i}_1^{n_i},z_\tau \in S_{t,\bullet},\theta)\\
\end{array}
\end{equation}  

In other words, each $\xi_i(t)$ corresponds to the mathematical expectation of the alignment times of the samples of the series $o_i$ with the sample $O_s(t)$ of the \textit{neutral} form. \\

In terms of algorithmic complexity, this algorithm is in $O(|\mathcal{O}|.||O_s||^2)$, that is to say quadratic with respect to the average size of the time series and therefore of the size $|O_s|$ of the \textit{neutral} form, and linear with respect to the cardinal of $\mathcal{O}$.

\begin{figure*}[]
\centering
\scalebox{.65}{
	\begin{tabular}{ccc}
		\includegraphics[width=90mm, height=80mm]{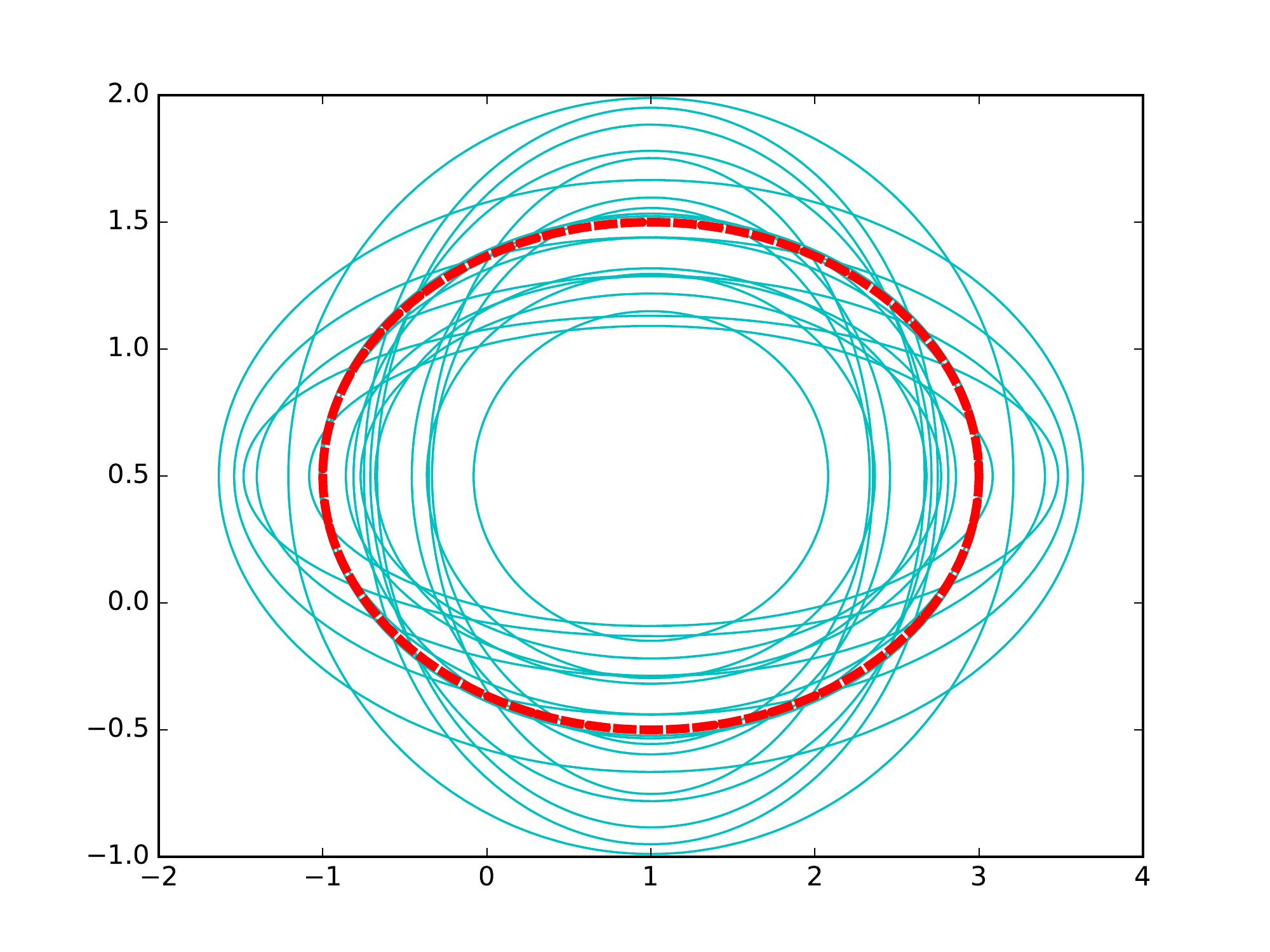} & \includegraphics[width=90mm, height=80mm]{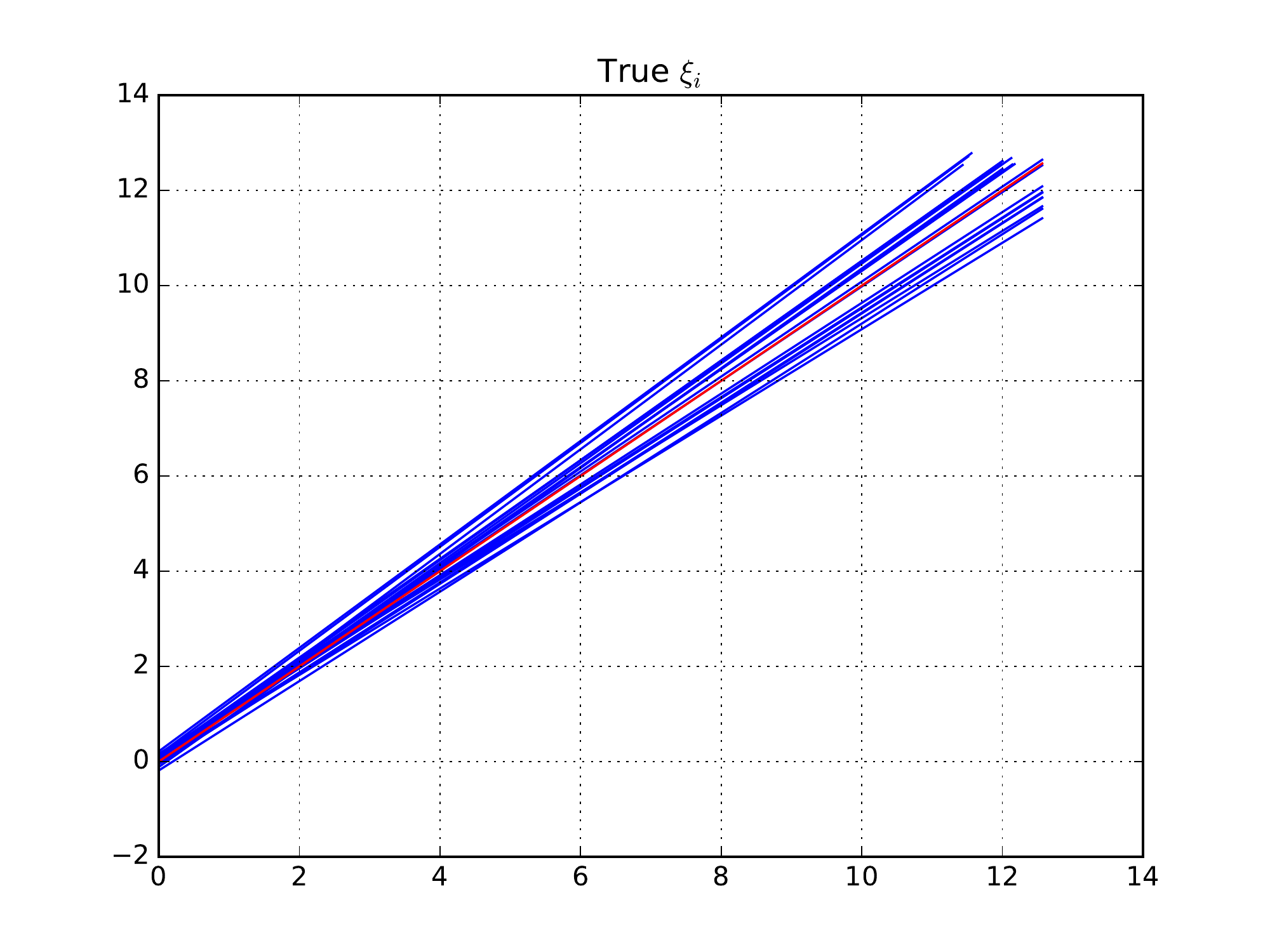} & \includegraphics[width=90mm, height=80mm]{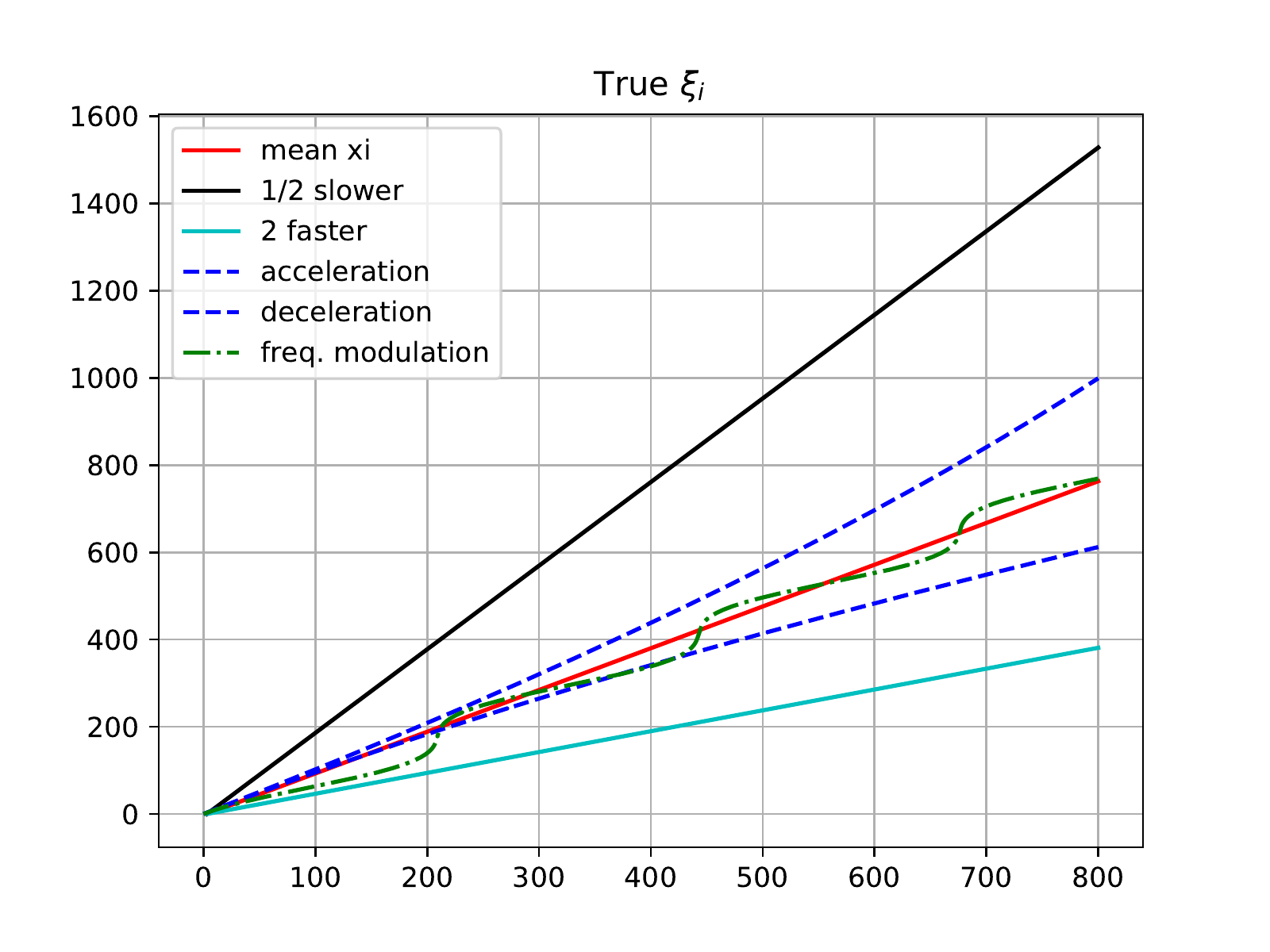}\\
		\includegraphics[width=90mm, height=80mm]{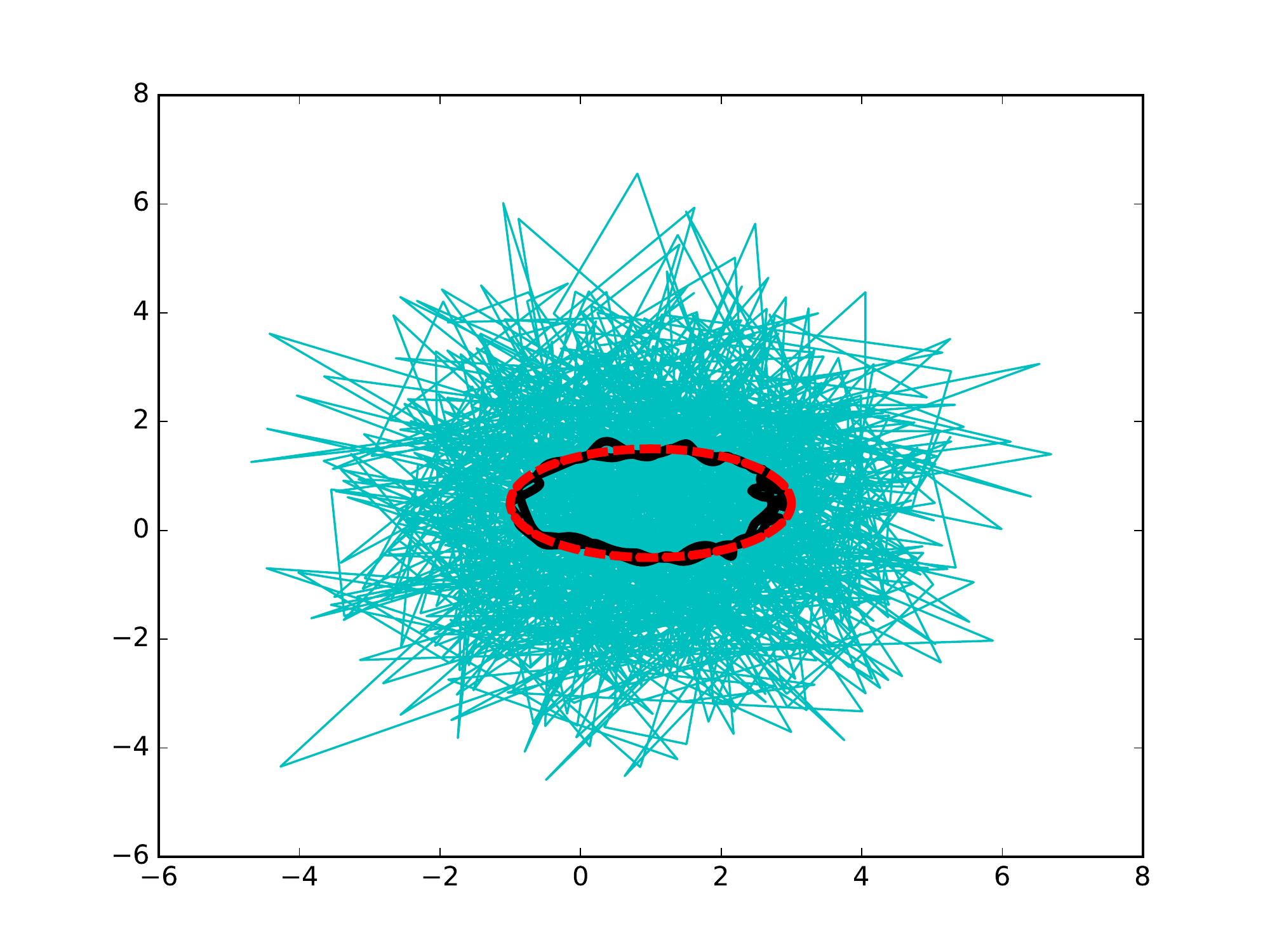}
		 	&
		\includegraphics[width=90mm, height=80mm]{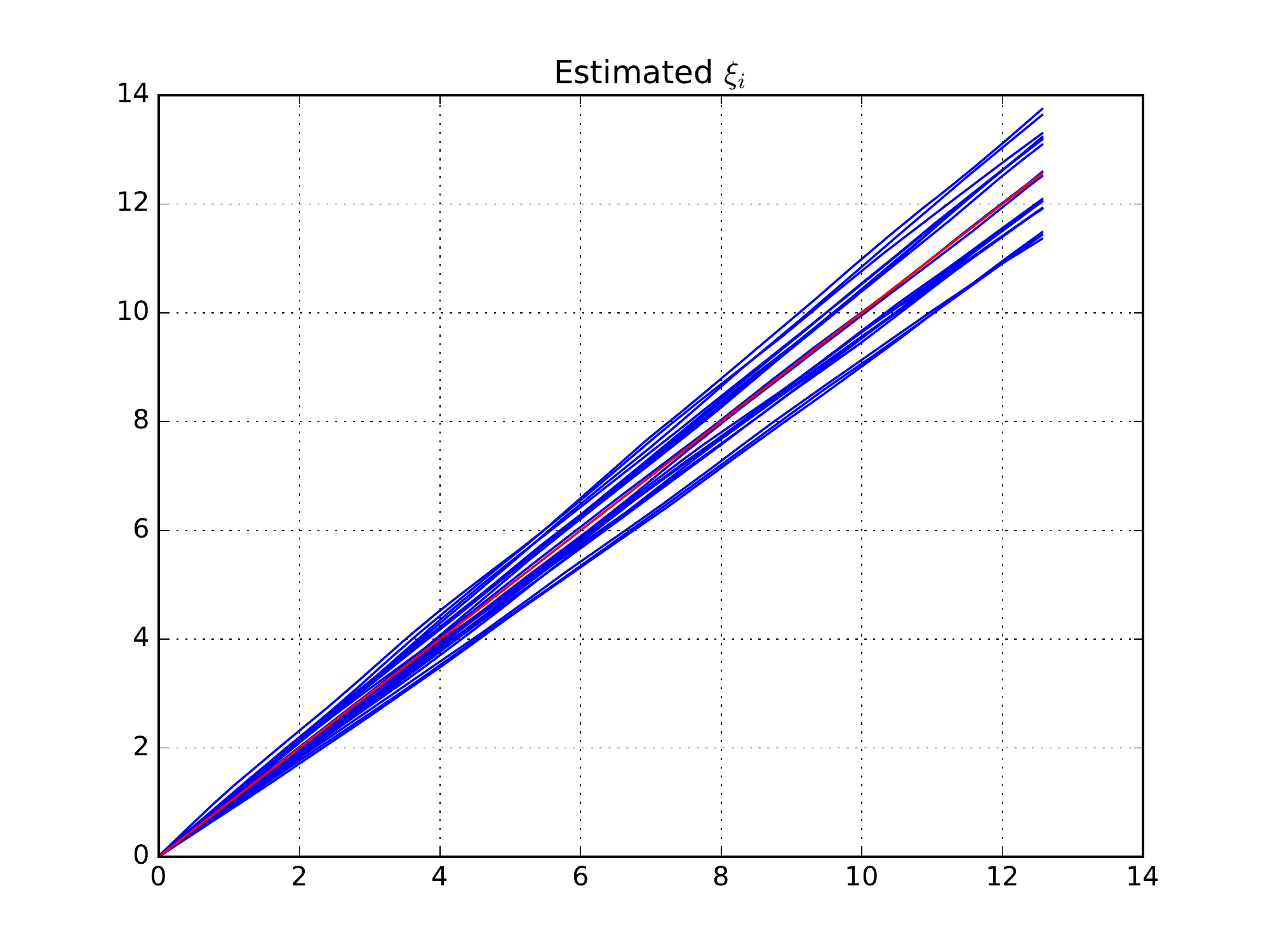} 
		 &
		\includegraphics[width=90mm, height=80mm]{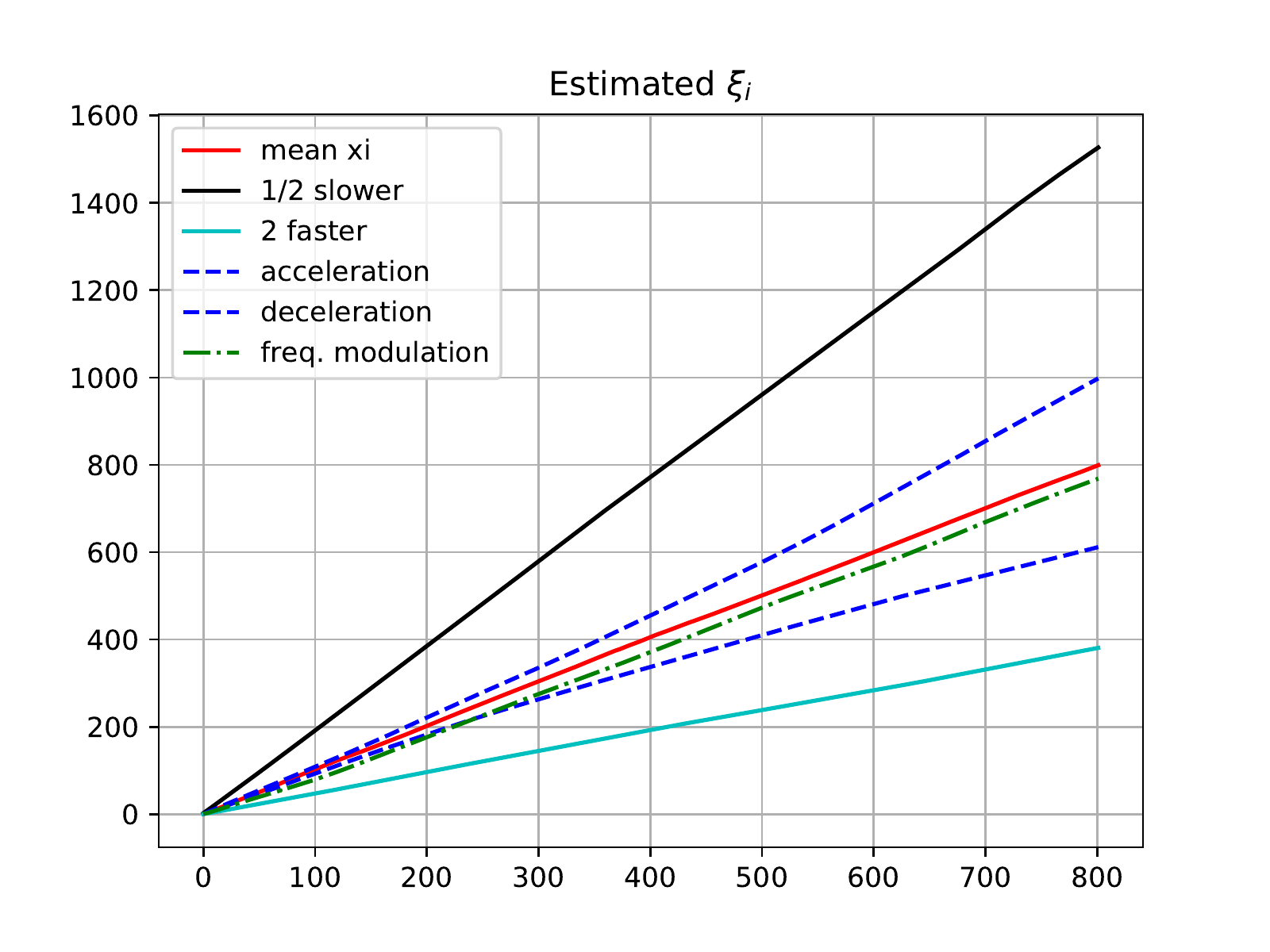}
	\end{tabular}}
	\caption{Top left sub-figure shows the variability of the 20 non-noisy ellipse shapes (theoretical \textit{neutral} shape in red). Bottom left sub-figure presents the 20 noisy time series of references, the exact \textit{neutral} shape (red curve and dashes), and the estimated \textit{neutral} shape $O_s$ (black curve). Top midlle sub-figure shows the (true) temporal variability of the temporal functions $\xi_i$ (in red $\xi_s$). Bottom middle sub-figure gives the estimated temporal functions $\xi_i$ (in red $\xi_s$). Top right sub-figure corresponds to the true $\xi_i$ functions associated to new noisy ellipse time series that have not been used to estimate the \textit{neutral} shape. Bottom right sub-figure is the estimation of the $\xi_i$ functions associated to these noisy ellipses. For all sub-figures, the function $\xi_s$ associated to the \textit{neutral} shape is shown in red color.}
	\label{fig:ELLIPSE}
\end{figure*}

\section{Experimentations}
We detail hereinafter two experiments in order to empirically support conjecture and to illustrate how the algorithm operates. The meta parameter $\nu$ of the local kernel is automatically optimized on the learning data, such as avoiding the convergences to zero of the calculated probabilities.

\subsection{Synthetic dataset}
We take again the ellipse example, previously introduced in subsection \ref{sec:context}, and characterized by the process $g \in \mathcal{G}$ defined through equation \ref{eq:ellipse_theta} with the following parameter values: $\omega_0=2\pi f_0$ with $f_0=1hz$ and $fe=400hz$ the sampling frequency, $\varphi_0=0$, $a_0=2$, $b_0=1$, all the ellipses being centered in $(1,.5)$, $\epsilon$ a Gaussian noise with standard deviation $\sigma=1.5$. 

A subset of $20$ time series is produced by $g$ from which the separation of shape and time functions is evaluated. The results of the separation process are shown in Fig. \ref{fig:ELLIPSE}.

To illustrate the variability of (non-noisy) shapes produced by $g$, the sub-figure at the top left shows the projections of the 20 ellipses randomly generated by $g$. At the bottom right sub-figure, are presented i) the noisy trajectories of the time series produced by $g$ (cyan curves), the theoretical \textit{neutral} (mean) shape (dashed red curve) and the \textit{neutral} shape estimate (black curve) provided by our algorithm. 

At the top center, the sub-figure presents the temporal variability by displaying the 20 true  $\xi_i$ temporal functions associated with the previous shapes.
The sub-figure at the bottom center shows the estimated temporal functions $\xi_i$, the red curve corresponding to the mean time curve associated with the \textit{neutral} shape.

Finally, the right sub-figures present new temporal functions produced by $g$ that violate the temporal constraints previously fixed. The true temporal functions are shown in the top sub-figure, while the estimated temporal function are shown in the bottom sub-figure. The red curve corresponds to the \textit{neutral} temporal curve $\xi_s$ associated to the \textit{neutral} shape $O_s$ ($ f0 = 1hz $). The cyan color curve corresponds to the estimate of a function $\xi$ generated with a frequency $f$ that is twice higher than $f_0$  ($f = 2hz$). The shape is thus traveled twice as fast and the slope expected for the cyan curve is twice times lower than that of the red curve, which is approximately the case for the estimation displayed in the figure. The black curve corresponds to the estimate of a function $\xi$ generated with a frequency $f$ twice as small as $f0$ ($f = 0,5hz$). The shape is therefore traveled half as fast as the \textit{neutral} shape, and the expected slope for the black curve is twice times greater than that of the red curve, which is approximately the case for the estimation presented in the figure. The blue dashed curves correspond to temporal curves presenting constant acceleration (below the red curve) and deceleration (above the red curve). The shapes of these two curves are no longer linear, which is also expected. Finally, the dashed green curve corresponds to a frequency modulated $xi$ curve, which thus presents successive accelerations and decelerations. The fluctuations are not very well reproduced in the estimation of this curve, but they are nevertheless quite visible in the  temporal curve that is extracted by our algorithm.

This simple synthetic example, which nevertheless presents a relatively important temporal and shape variability coupled with an additive noise largely covering the \textit{neutral} shape, shows that the initial objective behind this notion of separation of shape and time components within time series has some empirical grounds. The concept of time elastic centroid makes it possible, on this example, to approach this concept of \textit{neutral} shape and furthermore allows for the extraction of the temporal functions associated to the shape utterances that are randomly produced by a stochastic generative process $g$.

\subsection{On-line handwritten signature forgery detection}
We are interested here in some experiments related to the identification of forgeries of some online genuine signatures. These signatures are collected using capture devices such as digitizing tables, digital pens or touch screens. The data is in the form of trajectories in $\mathbb{R}^d$, with $d \ge 2 $ (including at least positions $x$ and $y$, and possibly pressure sensor output, stylus or pen orientations, etc.) 		

The tasks we consider are semi-supervised classification tasks that can be considered as being similar to anomaly detection tasks: only a few legitimate signatures are available for learning, forgery signatures can be considered as anomalies with respect to legitimate signatures.
		
The approach developed here consists in applying the separation algorithm on the training set, that is to say, to extract from legitimate signatures the \textit{neutral} shape component  and the temporal functions. From these shape and time components, we then construct an overall dissimilarity score between a test signature and our model that integrates a measure of shape dissimilarity and a measure of temporal dissimilarity.

The shape dissimilarity between a test time series $o$ and the \textit{neutral} shape $O_s$ is estimated by evaluating the mean squared error of the differences between the samples of the \textit{neutral} shape  $O_s(t)$ and the mathematical expectations of the samples of time series $o$ that are aligned with $O_s(t)$, as formalized by equation Eq. \ref{eq:dissimF}.
\begin{equation}
\delta_f(o, O_s)={\frac{1}{|O_s|}\sum_{t=1}^{|O_s|} \left(E(o|O_s(t))-O_s(t)\right)^2}
\label{eq:dissimF}
\end{equation}

The temporal dissimilarity is composed of the product of two terms, as specified by equation Eq. \ref{eq:dissimT}. The first term evaluates the absolute difference between the normalized averaged slope of temporal function $\xi_o$  and unity. The averaged slope is obtained by a linear regression estimated on the temporal function. It is normalized using the mean slope  $\overline{Slope}$ estimated on the linear regressions of the temporal function extracted from the training time series (which corresponds to legitimate signatures). 

The second term evaluates the absolute difference to unity of the area under the curve (\textit{AUC}) for temporal function $\xi_o$ normalized by the mean AUC, noted $\overline{AUC}$, of the $\xi$ functions extracted on the training data.
		
\begin{equation}
\delta_t(o, O_s)=(\frac{Slope(\xi_o)}{\overline{Slope}}-1)^2\cdot (\frac{AUC(\xi_o)}{\overline{AUC}}-1)^2
\label{eq:dissimT}
\end{equation}

The global dissimilarity score correspond to the fusion of the two previous terms as specified in equation Eq.\ref{eq:globalScore}. The meta parameter $\alpha$ is used to adjust weight of importance between temporal and shape dissimilarities. This method for separating shape and temporal functions in time series along with the subsequent score aggregation model is call STS (shape and time separation) in the remaining part of the article.

\begin{equation}
\delta(o, O_s)=\alpha\cdot log(1+\delta_f(o, O_s)) + (1-\alpha)\cdot log(\delta_t(o, O_s)+1)
\label{eq:globalScore}
\end{equation}

\begin{figure*}[!h]
	\centering
	\begin{tabular}{cc}
		\textbf{3D trajectories and \textit{neutral} shape} & \textbf{Temporal functions}, $\{\xi_i(t)\}$ \\
		\includegraphics[scale=.45]{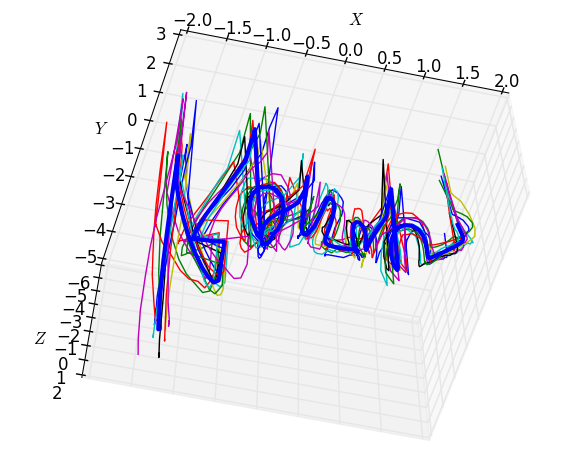}&
		\includegraphics[width=55mm, height=55mm]{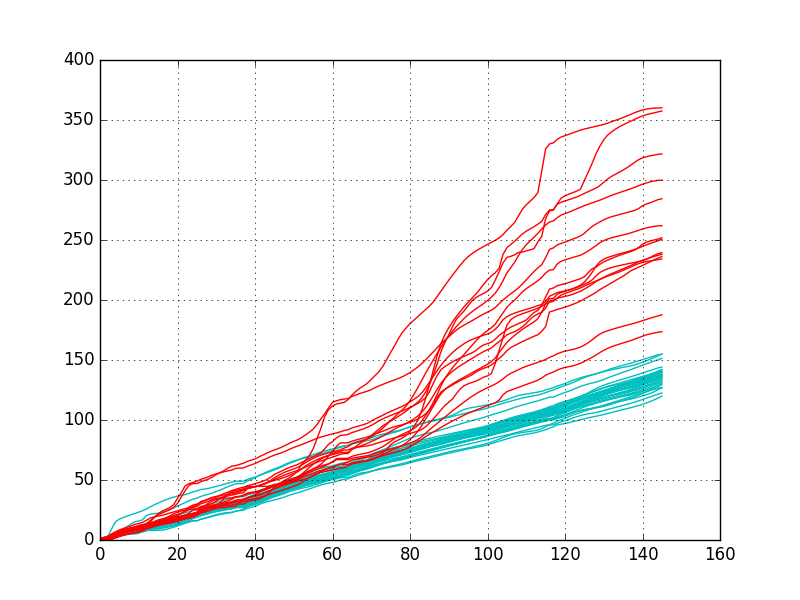} \\
		\textbf{2D score distributions}\\
		\includegraphics[scale=.27]{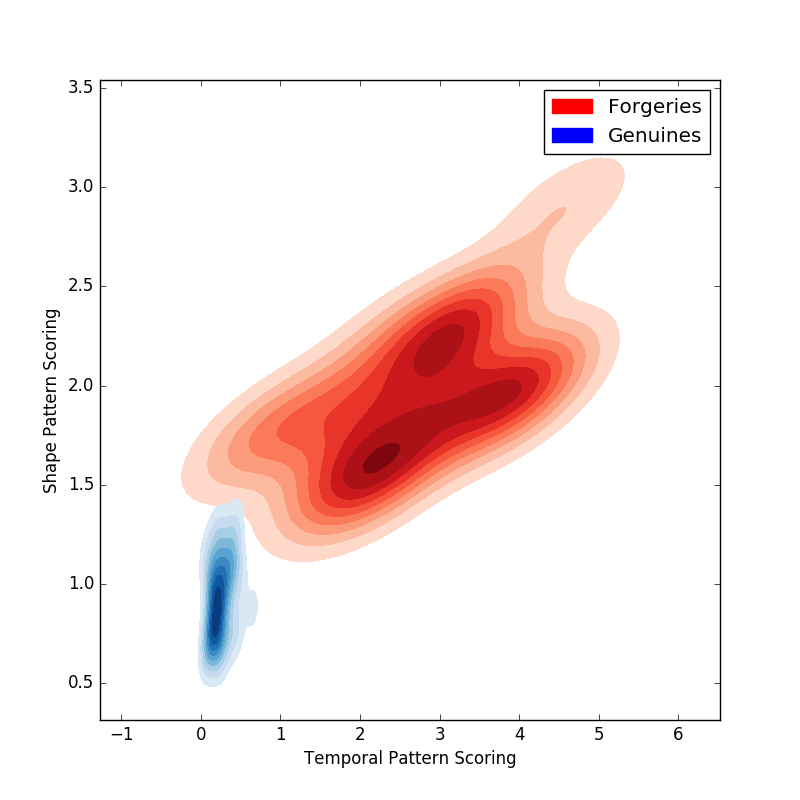}
	\end{tabular}
	\caption{Example from the MOBISIG dataset. Top left sub-figure: \textit{neutral} shape of a legitimate signature in dark blue, superimposed to the 3D trajectories from which it has been estimated. Top right sub-figure: temporal functions extracted from the legitimate signatures used for training (cyan colored curves) and from forgeries (red colored curves). Bottom sub-figure: 2D distributions of the dissimilarity scores ; x-axis the temporal dissimilarity scores, y-axis, the shape dissimilarity score. The cyan cluster corresponds to the scores of the legitimate signatures while the red one corresponds to the scores of the forgeries}
	\label{fig:MOBISIG}
\end{figure*}

As an illustration, we present in Fig. \ref{fig:MOBISIG}, the result of our separating algorithm applied on a case extracted from the  MOBISIG dataset presented below. On the top left sub-figure, the dark blue trajectory corresponds to the \textit{neutral} shape, superimposed to the curves that correspond to the legitimate signatures used for training. On the top right sub-figure, temporal functions extracted from legitimate signatures are shown in cyan color, while temporal functions extracted from forgery signatures are shown in red color. On this example, most of the forgeries are characterized with a highly nonlinear temporal function, whose slope greatly increases near the middle of the signature realization. Moreover, comparatively to genuine signatures, forgeries are produced with slower temporal patterns. Clearly, on this example, the average slopes and AUCs are significantly higher for forgeries than for genuine signatures.
		
In Fig. \ref{fig:MOBISIG}, bottom sub-figure,  we show the 2D distributions corresponding to the dissimilarity scores, with the temporal dissimilarity in the x-axis and the shape dissimilarity in the y-axis. One can notice that, for the legitimate signatures (blue, cyan cluster), this 2D distribution is much more compact than it is for the 2D distribution associated to forgeries (red cluster). Furthermore, on this example, using the two dissimilarity scores jointly enable us to find a better separating hyperplane between the two clusters comparatively to the case where only one of the two scores is considered.
		
For the considered \textit{benchmarks}, the assessment measures that are used are the following ones:
\begin{enumerate}
	\item EER : (\textit{equal error rate}) gives the error rate obtained when the false-positive rate is equal to the false-negative rate. In this configuration, rates are evaluated globally, i.e. when all subjects are treated together, without distinction.
	\item aEER : (\textit{equal error rate}) gives the average error rate obtained when the false-positive rate is equal to the false-negative rate. In this configuration, rates are evaluated subject by subject and then averaged.
	\item FAR et FRR : true False Acceptance Rate and False Rejection Rate obtained when operating at the closest to the EER point.\\
\end{enumerate}

The \textbf{SVC2004} \cite{Yeung2004svc2004} dataset is the first \textit{benchmark} specifically designed for evaluating  on-line handwritten signature verification systems. SVC2004 contains two distinct tasks: the first task only exploits the 2D (x,y) trajectories that are captured using a digitizing tablet (WACOM Intuos tablet) with a $100hz$ sampling frequency as well as the pen-down/pen-up binary information, that tells whether the contact between the pen and the tablet is established or not. The second task exploits supplementary information including the orientation of the pen. We have used the first task to determine a satisfactory value for meta parameter $\alpha$ (cf. Eq. \ref{eq:globalScore}). Using a cross-validation we have obtained best EER values in average for $\alpha=.85$. This $\alpha$ value has been used (and maintain constant) to evaluate the STS approach on the second MOBISIG task as well as for the processing of all the other benchmarks described below.\\

The \textbf{ICDAR2011} dataset \cite{Liwicki2011} is made of 3D trajectories $(x,y,z)$, produced by  10 subjects and captured using digitizing tablet  Intuos3 (with a sampling frequency of $200hz$) and a pen. Here $z$ contains the pressure exerted on the pen.\\

The \textbf{MOBISIG} dataset \cite{Antal2018} is made of pseudo-signatures produced by 83 subjects using one of their fingers and captured using a mobile tablet equipped with a touch screen.  The signatures are captured with a sampling frequency of $60hz$. The collected trajectories contain 2D positions $x$ and $y$, the speeds $vx$ and $vy$, accelerations $ax$ and $ay$ the pressure exerted on the touch screen  $z$, and the size of the contact area between the finger and the screen $ta$. We have applied the STS approach on the trajectories containing the previous features, without using the accelerations or the speeds dimensions that are more or less already encoded into the temporal functions we are extracting.\\

\begin{table*}[!h]
	\caption{Results obtained for each benchmarks. The evaluation measures are given in \%}
	\center
	\scalebox{1}{\begin{tabular}{c|c|c|c|c|c}
			\textbf{\textit{Benchmarks}} &	\textbf{Methods} & \textbf{aEER} & EER & FAR & FRR \\
			\hline\hline
			SVC2004 & EdlA,~\cite{Sharma2016} & 2.71 & - & - &-\\
			& \textbf{SFT} &\textbf{2.41} & 7.44 &  7.43 &7.45\\
			\hline
			ICDAR2011&	EdlA Dutch \cite{Liwicki2011} & - &- & 3.44 & 3.86 \\
			& \textbf{SFT Dutch} &  0.77 & 3.18 & \textbf{3.09}  & \textbf{3.27}\\
			&EdlA Chinese \cite{Liwicki2011} & -&-& 6.94 & 6.40 \\				
			&\textbf{SFT Chinese} &1.12 &3.44  & \textbf{3.69} & \textbf{3.20} \\				
			\hline
			MOBISIG (2017) & EdlA meilleur EER  \cite{Antal2018}&9.35  & 14.31 & - &-\\
			&EdlA meilleur aEER \cite{Antal2018}& 5.81  & 29.76  & - &-\\
			&\textbf{SFT} &\textbf{4.19} & \textbf{5.86} & 5.86 & 5.86\\
			\hline
			
	\end{tabular}}
	\label{tab:resultOLHWS}
\end{table*}
	
The pre-processing that has been implemented is common to all the datasets. Basically we remove the mean and normalized the data using standard deviations for $x$, $y$ and $z$ (the pen pressure). The other dimensions, if any, are normalized into the unit interval using $u=(u-min_u)/(max_u -min_u)$.

Table Tab. \ref{tab:resultOLHWS} presents in a synthetic way results that have been obtained on the three \textit{benchmarks} that cover four distinct tasks. Our algorithm (STS) achieves slightly better results than the best methods that have been evaluated so far on these tasks.  These results are quite  encouraging since the nature of the capture devices as well as the cultural differences of the subjects (European and Asian) are relatively diversified. The state of the art is mostly composed with methods that are derived from DTW for SVC2004 and MOBISIG benchmarks, and with statistical methods or heuristics-based approaches that are not detailed for the ICDAR2011 benchmark. Neural-based approaches, particularly those involving deep learning or recurrent architectures, seem poorly suited to these tasks because of the limited size of the training sets. \\

The C++ code with a Python wraper is available at\\ \url{https://github.com/pfmarteau/ShapeTimeSeparation/}.

\section{Conclusion}

We have stated in this paper the problem of the separation of shape and time components in time series. If, in its general formulation, this problem is under-constrained (\textit{ill-posed}), we have shown empirically, on synthetic data consisting of noisy ellipses, that a heuristic approach constructed around the notion of time elastic centroid can provide empirical solutions to this problem. In particular, the proposed algorithm makes it possible to identify a temporaly \textit{neutral} shape and a set of associated temporal functions that are coherent with respect to the time series that are processed and close to our initial expectations.

Furthermore, we have also evaluated the interest of this separation in the scope of a signature authentication task consisting in detecting forgeries from the knowledge of a small set of genuine signatures. On three benchmarks implementing capture devices of variable nature and precision, the proposed approach, based on an aggregation of temporal and shape dissimilarity scores,  slightly improves the results obtained by the best methods of the state of the art for each one of these benchmarks. These results are encouraging as the dissimilarity scores aggregating model that has been implemented is somehow quite simple.
 
The perspectives of this work concern the study and the modeling of the shape and temporal function variabilities in time series for the purposes of recognition or synthesis of shapes. The application aims are potentially very diverse, we can mention for instance the generation of expressive movements or realistic data that could be used to densify training sets.

Moreover, the presented algorithm is based on a global alignment principle, which is not adapted to the processing of certain class of time series such as chaotic series (because of the exponential dependence on the initial conditions). For such series, a solution based on temporally and spatially localized alignments could be possibly undertaken.

\normalsize

\ifCLASSOPTIONcompsoc
  \section*{Acknowledgments}
\else
  \section*{Acknowledgment}
\fi

The authors would like to thank...

\ifCLASSOPTIONcaptionsoff
  \newpage
\fi




\bibliography{biblio.bib}{}
\bibliographystyle{IEEEtran}
\bibliographystyle{plain}
%








\end{document}